\documentstyle[preprint,epsf,prl,aps]{revtex}
\begin{document}
\preprint{IUCM95-023}
\draft
\title{Is there an exponentially large Kondo screening cloud ?}
\author{Erik S.\ S\o rensen}
\address{Department~of~Physics,
Indiana~University, Bloomington, IN~47405}
\author{Ian~Affleck}
\address{Department of Physics and Canadian Institute
for Advanced Research}
\address{
University of British Columbia, Vancouver, BC, V6T 1Z1, Canada}
\date{August 9, 1995}
\maketitle
\begin{abstract}
We make a precise scaling conjecture, based on renormalization group
ideas, regarding the screening cloud around an impurity spin in the Kondo
effect and test it numerically using the Density Matrix Renormalization
Group method.
\end{abstract}

The renormalization group (RG) theory of the Kondo effect\cite{nozieres}
suggests the existence of an exponentially large length scale,
$\xi_K\equiv v_F/T_K \approx ae^{1/\rho J}$ where $v_F$ is the Fermi
velocity, $T_K$ the Kondo temperature, $a$  the unit cell dimension, $J$
the Kondo coupling and $\rho$ the density of states (per spin).  One
imagines a  cloud of electrons of this size, typically microns, which
screens the impurity spin.  The meaning, or even the existence of this
screening cloud has been a subject of some
controversy~\cite{slichter,lowD,Gan} with no clear consensus emerging
either experimentally or theoretically.  In particular, the experiments of
Boyce and Slichter\cite{slichter} were interpreted as indicating that the
screening cloud had a size of order a lattice spacing, about a hundred times
smaller than predicted by the renormalization group theory. In this letter we
make a more precise statement of what the existence of this screening cloud
really means, based on renormalization
group scaling ideas, and test it against
results from perturbation theory, local Fermi liquid theory and numerical
simulations.  A detailed account of our results will be presented
elsewhere~\cite{us}.
Related theoretical work includes perturbative calculations
\cite{Bressmann,Gan}, RG approaches \cite{Chen,Gan} and numerical work on the
Anderson model \cite{Gubernatis}. For a general review of the Kondo
effect see Ref.~\onlinecite{other} and references therein.

We consider the standard Kondo model:
\begin{equation}
H =
\sum_{\bf k}\epsilon_k \psi^{\dagger \alpha}_{\bf  k}\psi_{{\bf k}
\alpha} + J{\bf S}_{{\rm imp}}\cdot \sum_{\bf  k,\bf  k'}\psi^{\dagger
\alpha}_{\bf  k} \frac{\bbox{\sigma}^{\beta}_{\alpha}}{2}\psi_{{\bf  k'}
\beta},
\label{eq:hkk}
\end{equation}
with $S_{\rm imp}=1/2$.
We are interested in the Knight shift,
proportional to the local susceptibility:
\begin{equation}
\chi (r, T)
\equiv (1/T)<\psi^\dagger ({\bf r}) {\sigma^z\over 2}\psi ({\bf r})
S^z_{{\rm tot}}>.
\label{eq:locsus}
\end{equation}
Here $\bf S_{\rm tot}$ is the total spin operator, including both
impurity and conduction electron spins; we set the two g-factors equal.
In the scaling limit, $rk_F\gg 1$, $T\ll E_F$,
we conjecture the scaling form:
\begin{equation}
\chi (r,T,J)-\frac{\rho}{2} = {1\over v_Fr^2}\left\{
f\left({rT\over v_F}, {T\over T_K}\right)\cos \left[2k_Fr + \delta
\left({rT\over v_F}, {T\over T_K}\right)\right]+ g \left({rT\over v_F},
{T\over T_K}\right)\right\},
\label{scaling}
\end{equation}
where $f$, $g$ and $\delta$ are universal functions of two
scaling variables, and the bulk Pauli susceptibility has been subtracted.
The prefactor of $1/r^2$ arises from reducing the 3 dimensional problem to
an effective one-dimensional problem,\cite{Affleck} which is assumed to obey
scaling.
Instead of the second scaling variable, $T/T_K$, we may equivalently
use the renormalized Kondo coupling at scale $T$, $\lambda_{\rm eff}(T)$,
where $\lambda$ is the dimensionless Kondo coupling, $\lambda \equiv \rho
J$. For weak bare coupling and high temperatures, $T/T_K\approx \exp
[1/\lambda_{\rm eff}(T)]$.   Eq. (\ref{scaling}) follows naturally from the
expected asymptotic limits of $\chi$ and standard scaling hypotheses.  This
scaling form is probably the best definition of what it means to have a
screening cloud since it implies that for $T\le T_K, \chi(r)$ varies over at
least a distance of $\xi_K$.

Eq. (\ref{scaling}) is consistent with perturbation theory at  $T\gg T_K$,
where $\lambda_{\rm eff}(T)$ is small.  The first order result is:
\begin{equation}
\chi (r,T)
- \frac{\rho}{2} = {\lambda \over 16r^2v_F\sinh {2\pi rT\over  v_F}}\cos
2k_Fr,\label{RKKY}
\end{equation}
consistent with Eq. (\ref{scaling}).
Including the next order term,\cite{Bressmann,Gan} and assuming
also, $rT/v_F\ll 1$ we find:
\begin{equation}
\chi (r) -\frac{\rho}{2} =
- {\cos 2k_Fr\over 32\pi r^3T} \{\lambda +\lambda^2
[\ln (k_Fr)+{\rm constant}]\}.
\label{chiren}
\end{equation}
In low order perturbation theory,
the effective coupling at temperature $T$ is
given by\cite{nozieres}
\begin{equation}
\lambda_{\rm eff}(T) = \lambda + \lambda^2 \ln (v_Fk_F/T)+
O(\lambda^3).
\end{equation}
Thus, to $O(\lambda^2)$, the quantity in brackets in
Eq. (\ref{chiren}) may be written, as
\begin{equation}
\{ \lambda_{\rm eff}(T)
+ \lambda_{\rm eff}(T)^2[\ln (rT/v_F) + {\rm constant}]\},
\end{equation}
consistent with Eq. (\ref{scaling}).

Conversely, at $T\ll T_K$ and $r\gg \xi_K$, we expect local Fermi
liquid theory~\cite{nozieres} to apply.  $\chi$ then reduces to the magnetic
susceptibility of a non-magnetic impurity with a $\pi /2$ phase shift.
This can be obtained by differentiating
the Friedel oscillation formula for the
local charge density:
\begin{equation}
n(r) = n_0 -{1\over 2\pi^2r^3}\cos [2k_Fr+\pi
/2],
\end{equation}
with respect to the chemical potential yielding:
\begin{equation}
\chi
(r,T)={1\over 4v_F}{dn\over dk_F}={\rho \over 2} + {1\over
4\pi^2v_Fr^2}\cos (2k_Fr),
\label{chilt}
\end{equation}
again consistent with Eq. (\ref{scaling}).
Note that $\chi
(r,T)$ is longer-range at low T after the screening cloud has formed,
$\chi \propto 1/r^2$, than at higher T before it has formed, $\chi
\propto 1/r^3$, (Eq.~(\ref{chiren})).  Integrating Eq. (\ref{scaling})
over $d^3\bf r$, to obtain the long-range contribution to the total
susceptibility, only $g$ contributes, giving (after a change of
variables):
\begin{equation}
\chi_{\rm long\ range} = {4\pi \over
T}\int_0^\infty dx g(x,T/T_K).
\end{equation}
This is consistent with
the low temperature Fermi liquid prediction\cite{nozieres} $\chi \approx
1/T_K$ provided that
$\int_0^\infty dx g(x,y) \propto y \ ({\rm for}\  y\ll 1)$.

The functions $f(x,y)$, $g(x,y)$, and $\delta(x,y)$,
in the regime $y\leq 1$, $x/y =
r/\xi_K\leq 1$ are of special interest.  They describe the interior of
the screening cloud at low T.  One might naively suppose that a small
$r\ll \xi_K$ would also cut off the renormalization of the effective
coupling so that deep inside the screening cloud we recover
weak-coupling behavior (for weak bare coupling) even at low $T$.
However, explicit RG calculations by Gan~\cite{Gan} show this not to be
the case.  He showed that a small r {\it does not} cut off the infrared
divergence of the Kondo coupling in perturbation theory.  Thus the
scaling functions are expected to be non-trivial in this region.

Now let us consider the experiments of Boyce and Slichter\cite{slichter}
on Fe doped Cu. They measured the Knight shift
from  different shells of Cu atoms within a few lattice spacings from
the impurities from T=300K down to well below  the
Kondo temperature of 29K.  They found the factorized form: $ \chi (r,T) =
f(r)/( T+T_K)$,with some rapidly varying function $f(r)$ (which, in fact,
changes sign over the small range of r considered).  Note that all
measurements are taken in the regime $rT/v_F\ll 1$, $rT_K/v_F\ll 1$.  In
fact the values of $r$ are so small that it is unclear whether the
scaling form of Eq. (\ref{scaling}) holds at all.  Assuming it does, we
may essentially consider $x\to 0$ in the scaling functions. Assuming
that, in this limit, $f$ dominates over $g$, and that $\delta$ is
essentially constant, we can obtain the above result if $f(x,y)$ exhibits
an approximately factorized form for $x/y\ll 1$ and all $y$:
$f(x,y) \approx f(x/y)/ (y+1)$.
Note that such factorization could not also occur at large r, $r\gg \xi_K$
if our assumed scaling and asymptotic behaviors are correct.  In this
region, the $1/r^3$ behavior of the $2k_F$ part at $T\gg T_K$ crosses over
to $1/r^2$ at $T\ll T_K$.  Thus, a conclusive test of the existence of the
screening cloud would probably require experiments which probe length
scales of $O(\xi_K$).

To test our scaling hypothesis using the density matrix
renormalization group (DMRG) method~\cite{white} we consider a tight-binding
model with open boundary conditions:
\begin{equation}
H= -t\sum_{i=1}^{L-1}\left(\psi^{\dagger \alpha}_i\psi_{i+1,\alpha}
+\psi^{\dagger \alpha}_{i+1}\psi_{i,\alpha}\right)+
J{\bf S}_{\rm imp}
\cdot\psi_1^{\dagger \alpha}\frac{\bbox{\sigma}^\beta_\alpha}
{2}\psi_1^\beta,
\label{eq:tbh}
\end{equation}
with the total number of sites, $L$, even.
We shall always take the hopping strength, $t$, to be unity,
and we only consider ground-state properties at half-filling.
We keep $m=128$ states and for optimal precision
we apply the finite lattice DMRG~\cite{white}
to each step in the infinite lattice DMGR method.
In the ground-state the total magnetization, $S^z_{\rm tot}=1/2$,
is a constant and we therefore consider the local magnetization,
$<S^z_j>$, instead of $\Delta \chi \equiv \chi(r,T,J)-\rho/2$.
Here $S^z_j$ is the z-component of the electron spin at site $j$.
With the DMRG method we can obtain results for finite systems,
$L\le 40-50$, at $T=0$.
Hence we need a finite-size form of Eq.~(\ref{scaling}). This can
be obtained by noting that since $k_F=\pi/2$ at half-filling,
$\delta$ can be eliminated from Eq.~(\ref{scaling}). Substituting
$rT/v_F\to r/L$ and $T/T_K\to \xi_K/L$ we get:
\begin{equation}
<S^z_j> = {1\over L}\left\{
\widetilde f\left({j\over L}, {L\over \xi_K}\right)(-1)^j+
\widetilde g \left({j\over L},
{L\over \xi_K}\right)\right\}.
\label{eq:magscaling}
\end{equation}
In writing this equation we have included a factor of $T/r^2$
in going from 3D to 1D and converting to the local magnetization.

We briefly mention a few useful results regarding the weak and
strong coupling behavior of Eq.~(\ref{eq:tbh}).
In the weak coupling limit, $J\ll t$, using perturbation theory,
we find for large $L$:
\begin{equation}
<S^z_j>
\to {(J/t) \over 4\pi j}(-1)^j.
\label{eq:1sus}
\end{equation}
In the strong coupling limit, $J\gg t$, it can be shown that
the on-site magnetization behaves in the following way:
\begin{equation}
<S^z_j>=\frac{1}{L}\sin^2\frac{\pi j}{2},\ j > 2.
\label{eq:oddsus}
\end{equation}
These results are consistent with the scaling form Eq.~(\ref{eq:magscaling})
if we note that in the weak coupling limit we can
substitute $J/t$ with the effective
coupling $1/\ln(\xi_K/L)$.

\begin{figure}
\centering
\epsfysize=12 cm
\leavevmode
\epsffile{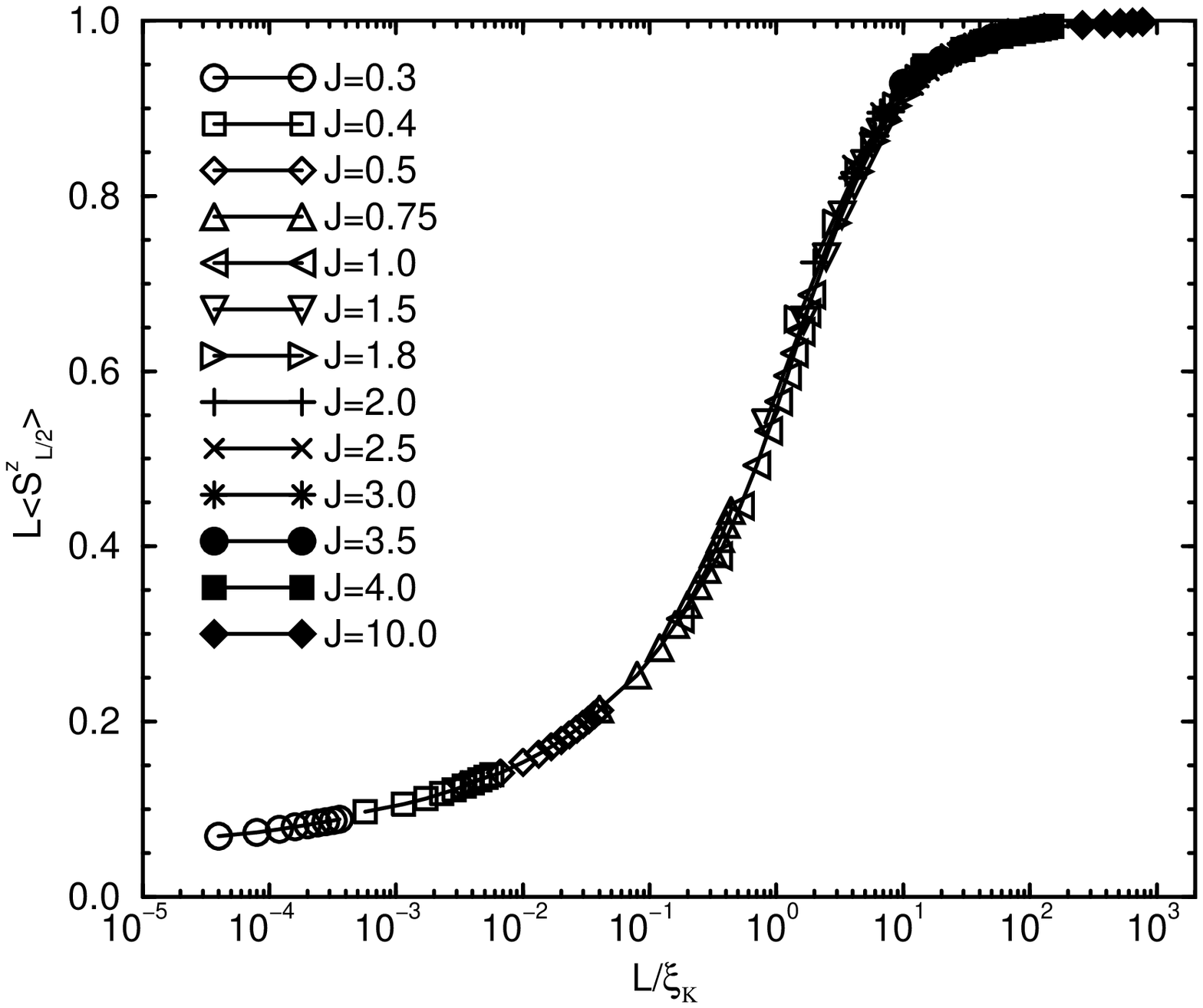}
\caption{$L<S^z_{L/2}>$ as a function of $L/\xi_K$ for
a range of different coupling constants.
The initial point corresponds in all cases to $L=4$, subsequent points
have L increased by 4.
}
\label{fig:l2scale}
\end{figure}

We start by considering $<S^z_{L/2}>$. Restricting our analysis
to {\it even} values of $L/2$ Eq.~(\ref{eq:magscaling}) takes the simpler
form $L<S^z_{L/2}>=h(L/\xi_K)$. If this scaling form is obeyed it should
be possible to collapse data for different $L$ and $J$ onto a single
scaling curve. In Fig.~\ref{fig:l2scale} we show our results for
a large range of Kondo couplings $J$ with $\xi_K(J=2.5)$ fixed
to 1. Choosing appropriate values
of $\xi_K$ all the data can clearly be brought onto the predicted scaling
form. A very dramatic crossover from weak to strong coupling behavior
is evident. Note that for strong coupling $L<S^z_{L/2}>$ approaches 1
consistent with the result Eq.~(\ref{eq:oddsus}). The weak coupling
result Eq.~(\ref{eq:1sus}) predicts a roughly constant value of
$<S^z_{L/2}>$ for fixed $J$
consistent with the behavior in Fig.~\ref{fig:l2scale}.
The fact that $<S^z_j> \to {\rm constant}/L$ for
$j\gg \xi_K$ is analogous to
the prediction that local susceptibility,
$\chi(r,T)$, should be longer range when
the screening cloud has formed than when it has not, as discussed above.

\begin{figure}
\centering
\epsfysize=12 cm
\leavevmode
\epsffile{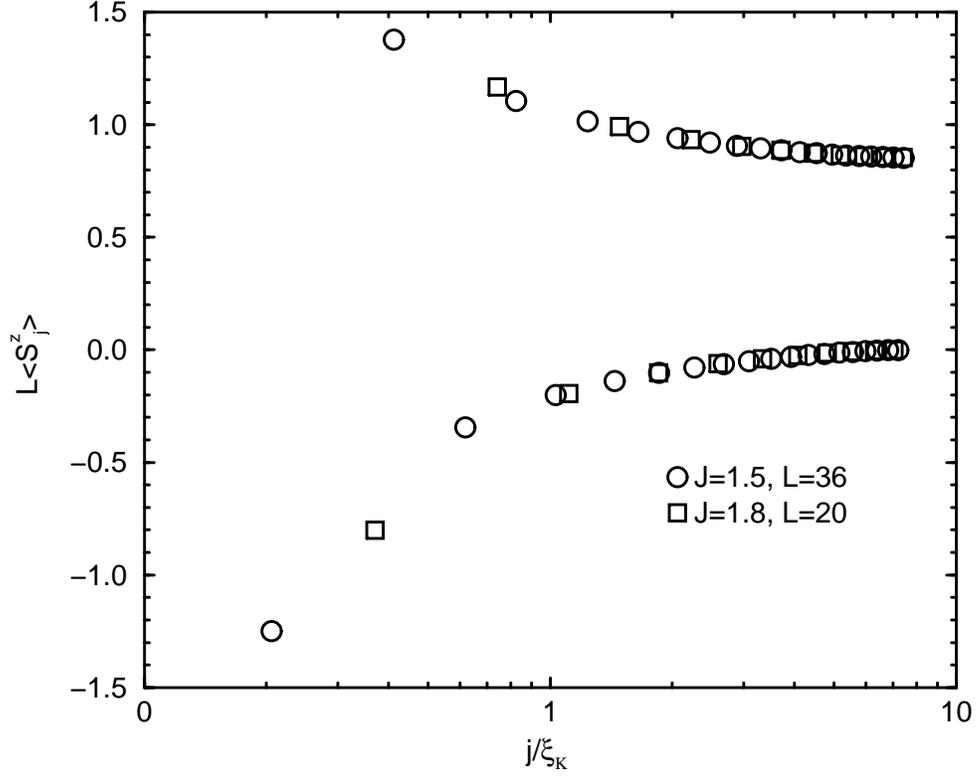}
\caption{$L$ times the expectation value of the z-component of the
electron spin, $<S^z_j>$, as a function of $j/\xi_K(J)$.
Two systems are shown:
$J=1.8,\ \xi_K=2.7,\ L=20$ and $J=1.5,\ \xi_K=4.85,\ L=36$.
Thus in both cases we have $L/\xi_K\approx 7.4$.
}
\label{fig:lsziscale2}
\end{figure}

We proceed to analyze the on-site magnetization
for fixed $L/\xi_K$ as a function of $j$. Our results are shown
in Fig.~(\ref{fig:lsziscale2}) where we display results for
$J=1.5$ and $J=1.8$ with $\xi_K(J=1.5)\sim 4.85$ and $\xi_K(J=1.8)\sim 2.7$,
and we use $L=36,20$ for $J=1.5,1.8$ respectively. Thus, $L/\xi_K\approx 7.4$
is kept fixed. An almost complete data collapse is clearly visible.
This is a highly non-trivial test of the scaling relation
Eq.~(\ref{eq:magscaling}) and hence also of Eq.~(\ref{scaling}).
Analogous scaling plots can be performed for the other couplings studied.

\begin{figure}
\centering
\epsfysize=12 cm
\leavevmode
\epsffile{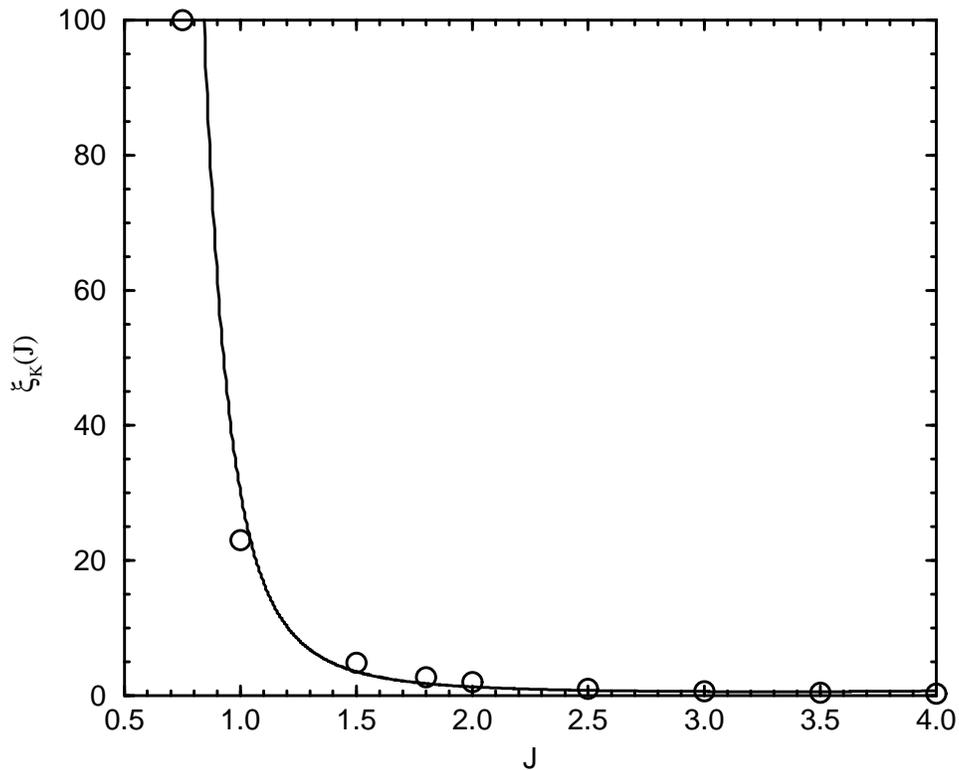}
\caption{The Kondo length, $\xi_K$, as a function of Kondo
coupling, $J$. The circles denote the numerical DMRG results.
The solid line indicates a least square fit of $\xi_K(J)$ for $J=1-4$
to the form Eq.~(\protect\ref{eq:xij}). The fitted parameters
are $\xi_0=1.76,\ c=0.21$.
}
\label{fig:xi}
\end{figure}

Having extracted $\xi_K$ from the scaling of $L<S^z_{L/2}>$,
we can study $\xi_K$ as a function of $J$. As
mentioned in the introduction, usual renormalization group
arguments predicts to lowest order an exponential form
for $\xi_K$.
Including one more term in the renormalization
group equation for the effective coupling we have~\cite{migdal},
\begin{equation}
-d\lambda /d\ln \Lambda = \lambda^2-\lambda^3/2.
\end{equation}
Here $\Lambda$ is the momentum space cut-off.
We can integrate this equation to get an improved expression for $\xi_K$.
This gives
\begin{equation}
\xi_K = \xi_0e^{1/\lambda_0-1/c}\sqrt{1-2/\lambda_0\over 1-2/c}.
\label{eq:xij}
\end{equation}
Here c is
positive and $O(1)$.
The inverse of the bare cut-off, $\xi_0$, should satisfy
$1 \ll \xi_0\ll \xi_K$ and $\lambda_0$, the bare coupling, is given by
$\lambda_0 = J\rho + (J\rho )^2\ln [\tan (2\xi_0/a)]$ with
the density of states for the tight-binding model, $\rho = 1/2\pi t$.
This form is valid for small $\lambda_0$. The corrections
arising from the constant $c$ is presumably only one of several equally
important terms; we have included it here to improve the agreement
with the numerical results. In the limit
$\lambda_0 \to 0$ Eq.~(\ref{eq:xij}) reduces to:
$\xi_K \propto e^{1/\lambda_0}/\sqrt{\lambda_0}$,
in agreement with previous results~\cite{zawa}.
The form, Eq.~(\ref{eq:xij}) is obtained using weak coupling RG
and should therefore
only be valid for the range of $J$ where $1 \ll \xi_K \ll L$.
In this region
we can now try to fit our previously obtained results for $\xi_K$
to the form Eq.~(\ref{eq:xij}). This is shown in Fig.~\ref{fig:xi}
where the DMRG results are shown as circles and the solid line
indicates a least square fit to Eq.~(\ref{eq:xij}) with the parameters
$\xi_0=1.76,\ c=0.21$.
{}For these intermediate couplings the
RG form works surprisingly well, and the data seems clearly consistent
with the expected form. However, one should bear in mind that
for weaker couplings $\xi_K$ has sizable error bars that we
cannot estimate; secondly, the extracted $\xi_K$ is only determined
up to a multiplicative constant.

In conclusion we have presented several non-trivial test
showing that the scaling form Eq.~(\ref{scaling})
clearly is obeyed in its finite-size form. Finite-size
effects may be important in recent experiments
using samples with lengths of order $\xi_K$ or smaller~\cite{lowD}.
The associated Kondo length scale, $\xi_K$, which roughly determines
the size of the screening cloud diverges at weaker
couplings with a behavior consistent with higher
order renormalization group calculations, Eq.~(\ref{eq:xij}).
The Knight shift associated with the local susceptibility is
longer range at low temperatures where the screening
cloud has formed than at higher temperatures where it has not.
Inside this screening cloud, as shown by Gan~\cite{Gan},
one does not expect weak coupling behavior to be recovered.

We would like to thank J.~Gan for many useful
and stimulating discussions.
We also thank W.~J.~L.~Buyers, S.~M.~Girvin,
B.~I.~Halperin, A.~H.~MacDonald, N.~Prokof'ev,
P.~C.~E.~Stamp, C.~Varma, E.~Wong and
A.~Zawadowski
for helpful comments. ESS is supported by NSF grant number NSF DMR-9416906.
This research was supported in part by NSERC of Canada.

\end{document}